\documentstyle[seg]{revtex}  
\begin{document}
\bibliographystyle{seg}

\title{
Kink Solution in a Fluid Model of Traffic Flows
}

\author{Shigeaki Wada\thanks{s-wada@yuragi.jinkan.kyoto-u.ac.jp} and Hisao Hayakawa\thanks{hisao@yuragi.jinkan.kyoto-u.ac.jp}}
\address{Graduate School of Human and Environmental Studies, Kyoto University, Kyoto 606-01}

\date{\today}

\maketitle

\begin{abstract}
Traffic jam in a fluid model of traffic flows proposed by Kerner and Konh\"auser (B. S. Kerner and P. Konh\"auser, Phys. Rev. {\bf{E 52}} (1995), 5574.) is analyzed.
An analytic scaling solution is presented near the critical point of the hetero-clinic bifurcation.
The validity of the solution has been confirmed from the comparison with the simulation of the model.
\end{abstract}

Cooperative behavior in dissipative systems composed of discrete elements has attracted much interest of physicists.
Traffic flows have been studied extensively~\cite{highway,old,old1,old2} as one of such systems.
They can be classified into discrete models and continuous models.
The former includes the cellular automaton model \cite{camodel} and the optimal velocity model\cite{bandou}, and the latter dose the fluid models.
Recently Kerner and Konh\"auser \cite{kkmodel} have proposed a fluid model of traffic flows which is regarded as a standard one.
Since these models are aimed to describe common phenomena such as traffic-jam formation, they should have universal properties as stressed by Hayakawa and Nakanishi~\cite{nishinomiya}.
The aim of this letter is to confirm the universality of traffic-jam formation.
For this purpose we quantitatively compare the results for the fluid model with those for a discrete model~\cite{hayanaka}.

Komatsu and Sasa \cite{komasasa} have revealed the mechanism of jam formation in the optimal velocity model proposed by Bando {\it{et al.}} \cite{bandou} with the aid of the perturbational treatment of solitons.
Hayakawa and Nakanishi \cite{hayanaka,nishinomiya} have developed the analysis of Komatsu and Sasa \cite{komasasa}, and analyzed in detail a generalized optimal velocity model
\begin{equation}\label{eq:ovmodel}
        \ddot x_n = a\bigl[W(x_{n+1} - x_n)\,V(x_n - x_{n-1}) - \dot x_n\bigr],
\end{equation}
where $x_n$ and $a$ are the positions of $n$-th car and the sensitivity, respectively.
The driver at $x_n$ takes care of not only the forward distance $x_{n+1} - x_n$ but also the backward distance $x_n - x_{n-1}$.
The optimal velocity function W is a monotonic increasing function, and V is a monotonic decreasing function.
Hayakawa and Nakanishi \cite{hayanaka,nishinomiya} have obtained a scaled asymmetric kink solution of this model.
On the other hand, Komatsu \cite{komatsu} has shown from his simulation that the fluid model by Kerner and Konh\"auser \cite{kkmodel} has also a scaled asymmetric kink solution as that in eq.~(\ref{eq:ovmodel}).
Here, we will clarify the relation between the model by Kerner and Konh\"auser and eq.~(\ref{eq:ovmodel}), based on the method developed by Hayakawa and Nakanishi \cite{hayanaka, nishinomiya}.
As a result, we will demonstrate that the behavior of the fluid model dose not contain essential differences from that of the discrete models.
In addition, the validity of our analysis will be confirmed from the comparison with our simulation of the model.

Kerner and Konh\"auser proposed a one-dimensional compressible fluid model for traffic flow in a highway
\begin{equation}\label{eq:kkmodel}
        \cases{
                \partial_t \phi = -\partial_z \phi v, \cr
                \partial_t v = -v \partial_z v + \displaystyle\frac{U(\phi) - v}{\tau}
                - \frac{T}{\phi} \partial_z \phi
                + \frac{\mu}{\phi}\partial_z^2 v,
        }
\end{equation}
where $\phi$ is the field of car density, and $v$ is the field of car velocity.
The term $(U(\phi) - v)/\tau$ with the relaxation time $\tau$ represents the relaxation process, where the optimal velocity field $U$ corresponds to $WV$ in eq.~(\ref{eq:ovmodel}).
The pressure term with the strength $T$ and the viscous term with the effective viscosity $\mu$ stabilize the solution of the model, which is the effects implicitly included in the model~(\ref{eq:ovmodel}).
The form of the optimal velocity $U(\phi)$ can be estimated from the relation between density and velocity of cars observed in real traffic flows in highways \cite{highway}.
Its explicit form is assumed to be
\begin{eqnarray}\label{U}
U(\phi) = 2.52305&\bigl[&\tanh\bigl((1-0.25)/0.12\bigr) \nonumber \\
& &{} - \tanh\bigl((\phi - 0.25)/0.12\bigr)\bigr],
\end{eqnarray}
as the paper by Kerner and Konh\"auser \cite{kkmodel}.

First, we consider the linear stability of a uniform flow, $(\bar{\phi},\bar{U} \equiv U(\bar{\phi}))$, in the fluid model~(\ref{eq:kkmodel}).
The linearized equation of the model~(\ref{eq:kkmodel}) in the neighborhood of this uniform state is
\begin{equation}\label{eq:linear}
        \cases{
                \partial_t \tilde{\phi} = -\bar{\phi}\partial_z \tilde{v}, \cr
                \partial_t \tilde{v} = \displaystyle\frac{U' \tilde{\phi} - \tilde{v}}{\tau}
                - \frac{1}{\bar\phi} (T \partial_z \tilde{\phi} - \mu \partial_z^2 \tilde{v}), \cr
        }
\end{equation}
where $\tilde{\phi} \equiv \phi - \bar{\phi}, \tilde{v} \equiv v - \bar{U}$$B!$(Band $U' \equiv d U/d \phi |_{\phi = \bar\phi}$.

The Fourier transform of eq.~(\ref{eq:linear}) satisfies
\begin{equation}\label{eq:senkeik}
\partial_t \pmatrix{\tilde{\phi} \cr \tilde{v} \cr}     =
        \pmatrix{
                0 & -i \bar\phi k \cr
                \displaystyle\frac{U'}{\tau}- i \frac{T}{\bar\phi} k & - \displaystyle\frac{1}{\tau} - \frac{\mu}{\bar\phi} k^2 \cr
        } \pmatrix{\tilde{\phi} \cr \tilde{v} \cr},
\end{equation}
where $k$ is a wave number.
The growth rate $\sigma$ in the solution $e^{\sigma t}$ of this equation is the solution of
\begin{equation}\label{eq:sigma}
        \sigma^2 + \left(\frac{1}{\tau} + \frac{\mu}{\bar\phi}k^2\right) \sigma
        + T k^2 + i \frac{\bar\phi U'}{\tau} k = 0.
\end{equation}
This equation has two solutions.
The condition for instability of the uniform state is $Re(\sigma_+)>0$ where $\sigma_+$ is one of the solution of eq.~(\ref{eq:sigma}),
\begin{equation}\label{eq:unstable1}
\frac{\bar\phi^2 U'^2}{\tau^2} k^2 > \left(\frac{1}{\tau} + \frac{\mu}{\bar\phi} k^2 \right)^2 T k^2.
\end{equation}
From eq.~(\ref{eq:unstable1}) we find that $k = 0$ is the neutral mode and the mode $k \to 0$ is the most unstable.
Thus, when
\begin{equation}\label{eq:unstable}
\bar\phi^2 U'^2 > T
\end{equation}
is satisfied, the uniform flow becomes unstable owing to emergence of the negative diffusion constant.

For later convenience, we explicitly write the expansion of $\sigma_+$ around $k = 0$.
\begin{eqnarray}\label{eq:longwave}
\sigma_+ &=& -i \bar\phi U' k + \tau (-T + \bar\phi^2 U'^2)k^2 \nonumber \\
& &     {} + i \tau U'(\mu - 2 \tau T \bar\phi + 2 \tau \bar\phi^3 U'^2)k^3 \nonumber \\
& & {} + \tau^2 \Bigl(- \tau T^2 + \frac{\mu T}{\bar\phi} - 3 \mu \bar\phi U'^2 \nonumber \\
& & {} + 6 \tau T \bar\phi^2 U'^2 - 5 \tau \bar\phi^4 U'^4 \Bigr)k^4 + \ldots.
\end{eqnarray}

Here, we notice that $\tau$ and $\mu$ are not included in the instability condition~(\ref{eq:unstable}). Then we assume $\tau = 1$ and $\mu = 1$ by a proper transformation of the scales of the space, time and field quantities.

Next, we carry out the weakly nonlinear analysis.
Firstly, let us reduce the model~(\ref{eq:kkmodel}) in the vicinity of neutral curve (\ref{eq:unstable}) with the aid of the reductive perturbation method \cite{reduct}.
From eq.~(\ref{eq:longwave}), the terms in proportion to $k^2$ and $k^4$, must be balance.
Furthermore, since the term $ik$ is eliminable by the Galilei transformation, the term $ik^3$ is the dominant in the dispersion relation, which should be balance with the lowest order nonlinear term. 
If the lowest nonlinear term is quadratic, the model is reduced to the Korteweg-de Vries (K-dV) equation \cite{hong,kdv,kdv1,kdv2,kdv3}.
The K-dV equation, however, has only pulse solutions, so it is unsuitable for a description of the traffic jam formation.
Therefore, we should choose the critical point which is the cross point of the neutral curve $T = \phi^2 U'(\phi)^2$ and $\,\,\phi U''(\phi) + 2U'(\phi) = 0$ where the cubic nonlinear term becomes dominant.
The explicit critical point when we adopt eq.~(\ref{U}) is given by
\begin{equation}\label{rinkai}
(\phi_c, T_c)=(0.300704, 28.2553).
\end{equation}

Now let us look for a steady propagating solution in the neighborhood of this critical point. 
We introduce the scaled variable $x$ 
\begin{equation}
        x \equiv \sqrt{\frac{c}{A}} \left(\epsilon (z - \sqrt{T}t) - c \epsilon^3 t \right),
\end{equation}
where $\epsilon \equiv \sqrt{(T_c - T)/T_c}, A = -U'$, and $c$ is the positive free parameter which will be determined from the perturbation analysis.
Since the lowest nonlinear term becomes cubic, we assume following expansions:
\begin{eqnarray}
\phi &=& \phi_c + \epsilon\,\sqrt{\frac{B}{c}}{\psi} + \epsilon^3\,{\phi_3}
        + \epsilon^4{\phi_4} + \ldots, \label{phi}\\
v &=& U(\phi_c) + \epsilon\,{v_1} + \epsilon^2\,{v_2} + \epsilon^3\,{v_3}
        + \epsilon^4\,{v_4}\ldots, \label{v} \hspace{0.5cm}
\end{eqnarray}
where $B = \phi_c U^{(3)}/6 - U'/\phi_c$ and $\psi$ includes $O(\epsilon)$.
We substitute these expression into eq.~(\ref{eq:kkmodel}) and collect the terms in each power of $\epsilon$.
From the expression to the fifth order of $\epsilon$, we obtain
\begin{equation}
\partial_x \left[ \partial_x^2 \psi - \psi (\psi^2 - 1)
        + \beta \partial_x \psi^2 \right]
        = \epsilon \partial_x M[\psi] \label{eq:reduce},
\end{equation}
where
\begin{eqnarray}
M[\psi] &=& \sqrt{c}\bigg[ \rho_{23} \partial_x^2 \psi^2
- \rho_{32} \partial_x \psi^3 \nonumber \\ & &\hspace{0.5cm}{}
- \rho_{41} \psi^4
+ \rho_{14} \partial_x^3 \psi
- \frac{\rho_{12}}{c} \partial_x \psi \bigg] \nonumber \\ & &{}
+ \bigg[ \frac{\dot c}{2 c^{\frac{5}{2}}}x \psi_0 - c \dot c t \psi \bigg], \label{M}
\end{eqnarray}
where we assume $\dot c$ is $O(\epsilon)$,
$
\beta = C/\sqrt{A B},\,
\rho_{23} = D/\sqrt{A^2 B},\,
\rho_{32} = E/\sqrt{A B^2},\,
\rho_{41} = F/\sqrt{B^3},\,
\rho_{14} = G/\sqrt{A^3},\,
C = \phi_c U'^2,\,
D = -2 \phi_c^2 U'^3 - U'/\phi_c,\,
E = -\phi_c^2 U' U^{(3)}/3 + U'^2,\,
F = -\phi_c U^{(4)}/24 - U^{(3)}/6,\,
G = 2 \phi_c U'^2,$
and
$
H = \phi_c^2 U'^2.
$

Here, assuming the expansion
\begin{equation}
\psi(x) = \psi_0(x) + \epsilon \psi_1(x) + \ldots,
\end{equation}
we obtain the following pair of asymmetric kink and anti-kink solutions in the lowest order.
\begin{equation}\label{eq:kink}
\psi^{(\pm)}_0(x) = \tanh(\theta_{\pm} x);\quad \theta_{\pm} = \frac{\beta \pm \sqrt{\beta^2+2}}{2}.
\end{equation}
Let us assume that the system satisfies the periodic boundary condition.
Thus the solution should be modified as
\begin{equation}\label{eq:zero-pbc}
\psi_0(x)\simeq {\psi_0}^{(+)}(x - x_+)-1+{\psi_0}^{(-)}(x - x_-).
\end{equation}
This solution shows two interfaces at the positions $x = x_\pm$, and approximately satisfies the periodic boundary condition, though we notice that eq.~(\ref{eq:zero-pbc}) is only an approximate solution in the lowest order equation of (\ref{eq:reduce}).
Substituting (\ref{eq:zero-pbc}) into (\ref{eq:reduce}), we obtain
\begin{equation}\label{eq:linearized}
        {\cal L} \psi_1 = \frac{d}{dx} M[\psi_0],
\end{equation}
where
\begin{eqnarray} \label{eq:L-op}
{\cal L} &=& \partial_x^3+\partial_x-6\psi_0\partial_x-3\psi_0^2\partial_x \nonumber \\
& & {} + 2\beta \partial_x^2\psi_0+4\beta \partial_x\psi_0\partial_x+2\beta \psi_0\partial_x^2.
\end{eqnarray}
To obtain a regular behavior of perturbation in $O(\epsilon)$ the perturbed solution should be orthogonal to the zero eigenfunction \cite{hayanaka}.
This solvability condition is given by
\begin{equation}\label{solva}
\bigl(\Psi_0,\partial_x M[\psi_0]\bigr)\equiv 
\lim_{L\to \infty}\int_{-L}^{L} dx \Psi_0 \partial_x M[\psi_0]=0,
\end{equation}
where $L$ is the system size, and $\Psi_0$ is a zero eigenfunction which satisfies 
\begin{equation}\label{zero-eq}
{\cal L}^{\dagger}\Psi_0=0 ; \quad 
{\cal L}^{\dagger}=-\partial_x^3-\partial_x+3 \psi_0^2\partial_x+
2\beta \psi_0\partial_x^2.
\end{equation}
Equation~(\ref{zero-eq}) can be rewritten as
\begin{equation}\label{phi0}
\tilde{\cal L}^{\dagger}\Phi_0(x)=0; \quad 
\tilde{\cal L}^{\dagger}=-\partial_x^2-1+3 \psi_0^2+
2\beta \psi_0\partial_x,
\end{equation}
where $\Phi_0 \equiv \partial_x \Psi_0$.
A pair of special solutions of eq.~(\ref{phi0}) is given by
\begin{equation}
\Phi_0^{(\pm)}(x)=\bigl({\rm sech}\,(\theta_{\pm} x)\bigr)^{1/\theta_{\pm}^2},
\end{equation}
and the corresponding solutions of (\ref{zero-eq}) can be expressed by 
\begin{equation}\label{psi}
{\Psi_0}^{(\pm)}(x)=\frac{\alpha_{\pm}}{2}\int_{-x}^xdx' 
\bigl({\rm sech}\,(\theta_{\pm} x')\bigr)^{1/\theta_{\pm}^2}.
\end{equation}
Since $\Psi_0$ should satisfy the periodic boundary condition as well as $\psi_0$, we adopt the superposition of $\Psi_0^{(\pm)}$ as
\begin{equation}\label{psi0}
\Psi_0(x)={\Psi_0}^{(+)}(x-x_+)-1+{\Psi_0}^{(-)}(x-x_-).
\end{equation}
The constants $\alpha_{\pm}$ in eq.~(\ref{psi}) are determined to satisfy $\Psi_0(\pm \infty) = -1$.
Thus, we choose
\begin{equation}\label{alpha}
\alpha_{\pm} = \frac{2\theta_{\pm}}{{I_0}^{(\pm)} },
\end{equation}
where
\begin{eqnarray}
 I_n^{(\pm)} &=& \int_{-\infty}^{\infty} dx ({\rm sech}\,x)^{1/\theta_{\pm}^2+2n} \nonumber \\
&=& \sqrt{\pi}
\frac{\Gamma\left(\frac{1}{2\theta_{\pm}^2}+n\right)}{\Gamma\left(\frac{1}{2\theta_{\pm}^2}+n+\frac{1}{2}\right)}
\end{eqnarray}
with the Gamma function $\Gamma (x)$.

Now, we rewrite the solvability condition~(\ref{solva}) by the integration by parts
\begin{equation}\label{solva2}
\big[\Psi_0 M[\psi_0] \big]_{-L}^{L}=\bigl(\Phi_0(x), M[\psi_0]\bigr).
\end{equation}
From (\ref{M}) it is obvious that terms in $M[\psi_0]$ have no contribution in the left hand side of (\ref{solva2}) except for terms in proportion to $\rho_{41}$ and $x\psi$.
Notice that the contribution from the term in proportion to $t\psi$ vanishes because of its symmetry.
In addition, since we adopt the periodic boundary condition, as in eqs.~(\ref{eq:zero-pbc}) and (\ref{psi0}), the contribution from the term $\rho_{41}$ is canceled.
Therefore the left hand side of (\ref{solva2}) is reduced to
\begin{equation}\label{b-eff}
\big[\Psi_0 M[\psi_0] \big]_{-L}^{L}=\frac{\dot c}{c^{\frac{5}{2}}}L.
\end{equation}
After computation of the right hand side of eq.~(\ref{solva2}), we obtain the time evolution equation of $c$,
\begin{eqnarray}\label{hatten}
\big[L-(\theta_+-\theta_-)\big]\dot c =4\beta c^2&\biggl[&
\frac{\theta_+}{\theta_+^2+1}\left(1-\frac{c}{c_+}\right) \nonumber \\
& &{}-\frac{\theta_-}{\theta_-^2+1}\left(1-\frac{c}{c_-}\right)\biggr],\hspace{0.5cm}
\end{eqnarray}
where we use
\begin{equation}
\frac{I_{n+1}^{(\pm)}}{I_{n}^{(\pm)}} = \frac{2n\theta^2_{\pm}+1}{(2n+1)\theta^2_{\pm}+1}.
\end{equation}
Here $c_{\pm}$ are given by
\begin{eqnarray}
c_{\pm}^{-1} &=& -2\theta_{\pm}^2 \frac{\rho_{14}}{\rho_{12}} \left( 2 - 3 \frac{I^{(\pm)}_2}{I^{(\pm)}_1} \right) \nonumber \\
& &{}+ 2 \theta_{\pm} \frac{\rho_{23}}{\rho_{12}}\left( 2 - 3 \frac{I^{(\pm)}_2}{I^{(\pm)}_1} \right) \nonumber \\
& &{} + 3 \frac{\rho_{32}}{\rho_{12}}\left( 1 - \frac{I^{(\pm)}_2}{I^{(\pm)}_1} \right) \nonumber \\
& &{} + \frac{\rho_{41}}{\theta_{\pm} \rho_{12}} \left( \frac{I^{(\pm)}_0}{I^{(\pm)}_1} -2 + \frac{I^{(\pm)}_2}{I^{(\pm)}_1} \right),
\end{eqnarray}
which are expected to be the stable values of $c$ where a kink or an anti-kink exists under the free boundary condition.
Equation~(\ref{hatten}) represents that the relaxation time for a steady state is divergent in proportion to the system size $L$ in the vicinity of the convergent value $c^*$ of c.
From eqs.~(\ref{U}) and (\ref{rinkai}), each constant is determined as
$
\beta = 2.01476,\,
\theta_+ = 2.23815,\,
\theta_- = -0.223398,\,
\rho_{23} = 5.39424,\,
\rho_{32} = 1.92455,\,
\rho_{41} = 0.299797,\,
\rho_{14} = 2.52857,\,
\rho_{12} = 6.72039
$.
Therefore we obtain
\begin{equation}
c_+ = 2.67997,\quad c_- = 2.62752.
\end{equation}
and $c^*$ is given by
\begin{equation}
c^* = \frac{c_+ c_- \bigl[\theta_+ (\theta_-^2 + 1) - \theta_- (\theta_+^2 + 1)\bigr]}{c_- \theta_+ (\theta_-^2 + 1) - c_+ \theta_- (\theta_+^2 + 1)}.
\end{equation}
Thus the explicit value of $c^*$ is
\begin{equation}\label{eq:cstar}
c^* = 2.66066.
\end{equation}

To check the validity of our analysis we perform the numerical simulation of the model~(\ref{eq:kkmodel}) in the vicinity of the critical point under the periodic boundary condition.
We adopt the classical fourth-order Runge-Kutta scheme to time and the Eular scheme to space in our simulation.
The analytic solution in eqs.~(\ref{eq:kink}), (\ref{eq:zero-pbc}), and (\ref{eq:cstar}) was adopted as the initial condition for the fast convergence.
We also have checked that the tendency to converged to $c^*$ from the random initial condition.
Taking into account the scaling properties we perform the simulation for each parameter, $\epsilon = 1/2, 1/4, 1/8,$ and $1/16$, until $\phi$ relaxes to a steady propagating state.
Here we change the space mesh size according to the divergence of the characteristic length in proportion to $\epsilon \to 0$.

\begin{figure}
        \caption{\small{The theoretical curve in eqs.~(\ref{eq:kink}), (\ref{eq:zero-pbc}), and (\ref{eq:cstar}) and data of the scaled density obtained by our simulation. The horizontal axis is displays the car position. The system size is normalized to 1, and positions of the interfaces are set to $0.25$ and $0.75$ which are the fitting parameters of the theoretical expression. Each data is obtained from the result for $\epsilon = 1/2, 1/4, 1/8$, and $1/16$.}}%
        \label{fig:kink}%
\end{figure}%
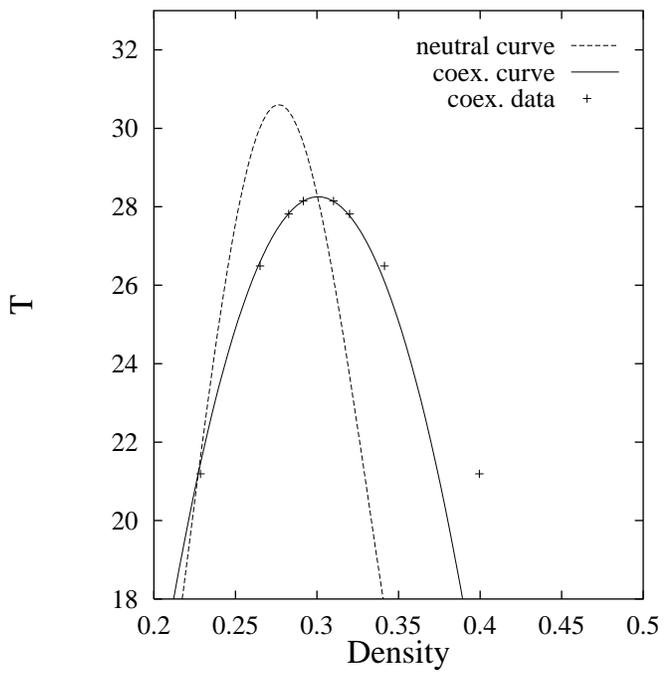
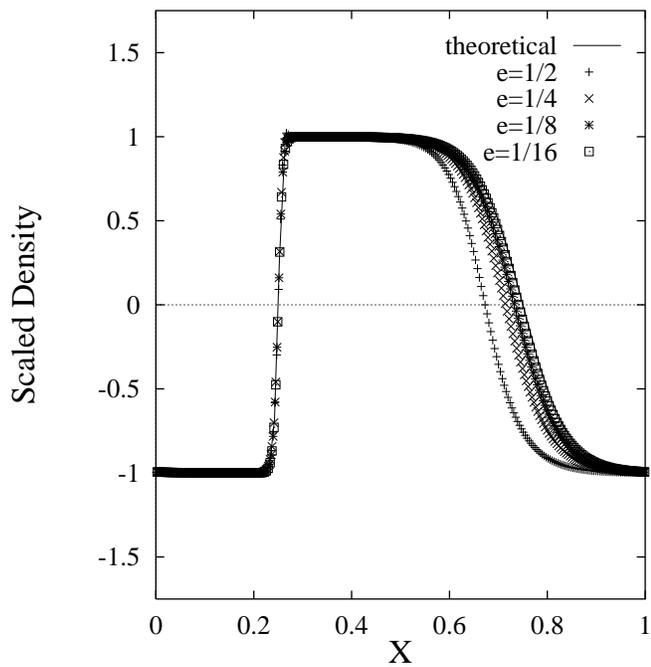

Our results are plotted in Figs.~\ref{fig:coex} and \ref{fig:kink}.
In Fig.~\ref{fig:coex}, points represent the maximum and minimum values of car density in each parameter. The solid curve represents the theoretical coexistence curve which is obtained from the above perturbation analysis as
\begin{equation}
T = T_c \left( 1 - \frac{B(\phi - \phi_c)^2}{c^*} \right).
\end{equation}
We emphasize that the deviation between simulation and theory is only $1.0\%$ at $\epsilon = 1/16$.
It is interesting that we can see one of the branches in the linearly unstable region.
Figure~\ref{fig:kink} displays the scaled density of cars as the function of the scaled position of cars. %
The solid line is theoretical one in eqs.~(\ref{eq:kink}), (\ref{eq:zero-pbc}), and (\ref{eq:cstar}).
The result of our simulation is asymptotically identical to our theoretical curve as $\epsilon\to 0$, and reproduce an asymmetric pair of  kink and anti-kink.
As we have seen in this letter, our results of the simulation agree well with our theoretical analysis.

In conclusion, our results in the fluid model in eq.~(\ref{eq:kkmodel}) can be summarized as follows:
\begin{enumerate}
        \item The scaling solution exists in the vicinity of the critical point.
        \item There is an asymmetry of the solution.
        \item Quantitative results can be reproduced by the perturbation analysis.
\end{enumerate}
These results correspond to those in the analysis of discrete model~(\ref{eq:ovmodel})\cite{hayanaka,nishinomiya}.
Thus we suggests that the results obtained here are universal ones which are independent of choice of a specific model.

\section{Acknowledgements}
One of the authors (HH) thanks K. Nakanishi for fruitful discussion. SW wishes to thank T. Takaai for helpful conversations. This study is supported by the Grant-in-Aid of Ministry of Education, Science and Culture of Japan (09740314).

\newpage
\begin{figure}
        \caption{\small{The theoretical coexistence curve (solid line) in eq.~(\ref{hatten}) and the neutral curve (dotted line) in eq.~(\ref{eq:unstable}) in $(\phi, T)$ plane. The data points (cross) are obtained from the maximum and minimum values of car density $\phi$ at given $T$.}}
        \label{fig:coex}
\end{figure}
\end{document}